\documentclass[letter,traditabstract]{aa}
\usepackage[dvips]{graphicx}
\usepackage{txfonts}
\def\kms{\mbox{km~s$^{-1}$}}

\def\Vlsr{$V_{\rm LSR}$}

\def\jdu{\mbox{2--1}}
\def\juz{\mbox{1--0}}
\def\jtd{\mbox{3--2}}

\begin{document}
\title{Close encounters of the protostellar kind in IC~1396N\thanks{Based 
on observations carried out with the IRAM Plateau de Bure. IRAM is supported by 
INSU/CNRS (France), MPG (Germany), and IGN (Spain).}} 
\author{M.\ T.\ Beltr\'an\inst{1} \and F.\ Massi \inst{1}  \and 
F.\ Fontani \inst{1} \and C.\ Codella \inst{1, 2} \and R.\ L\'opez \inst{3}}
\institute{
INAF, Osservatorio Astrofisico di Arcetri, Largo E.\ Fermi 5,
50125 Firenze, Italy
\and
UJF-Grenoble 1/CNRS-INSU, Institut de Plan\'etologie et d'Astrophysique de 
Grenoble (IPAG) UMR 5274, Grenoble, France
\and
Departament d'Astronomia i Meteorologia, Universitat de Barcelona, Mart\'{\i} i
Franqu\`es 1, 08028 Barcelona, Catalunya, Spain
}
\offprints{M.\ T.\ Beltr\'an, \email{mbeltran@arcetri.astro.it}}
\date{Received date; accepted date}

\titlerunning{Close encounters in IC~1396N}
\authorrunning{Beltr\'an et al.}

\abstract { 
We have mapped in the 2.7~mm continuum and $^{12}$CO with the PdBI
the IR-dark ``tail"  that crosses the IC~1396N globule from south to north,
and is the most extincted part of this cloud. These observations have
allowed us to distinguish all possible associations of molecular hydrogen
emission features by revealing two well-collimated low-mass protostellar 
outflows at the northern
part of the globule. The ouflows are located almost in the plane of the sky and
are colliding with
each other towards the position of a strong 2.12~$\mu$m H$_2$ line emission feature.
} 
\keywords{ISM: individual objects: IC~1396N -- ISM: jets and outflows -- stars: formation -- infrared: ISM}

\maketitle

\section{Introduction}

\begin{figure*}
\centerline{\includegraphics[angle=0,width=10cm]{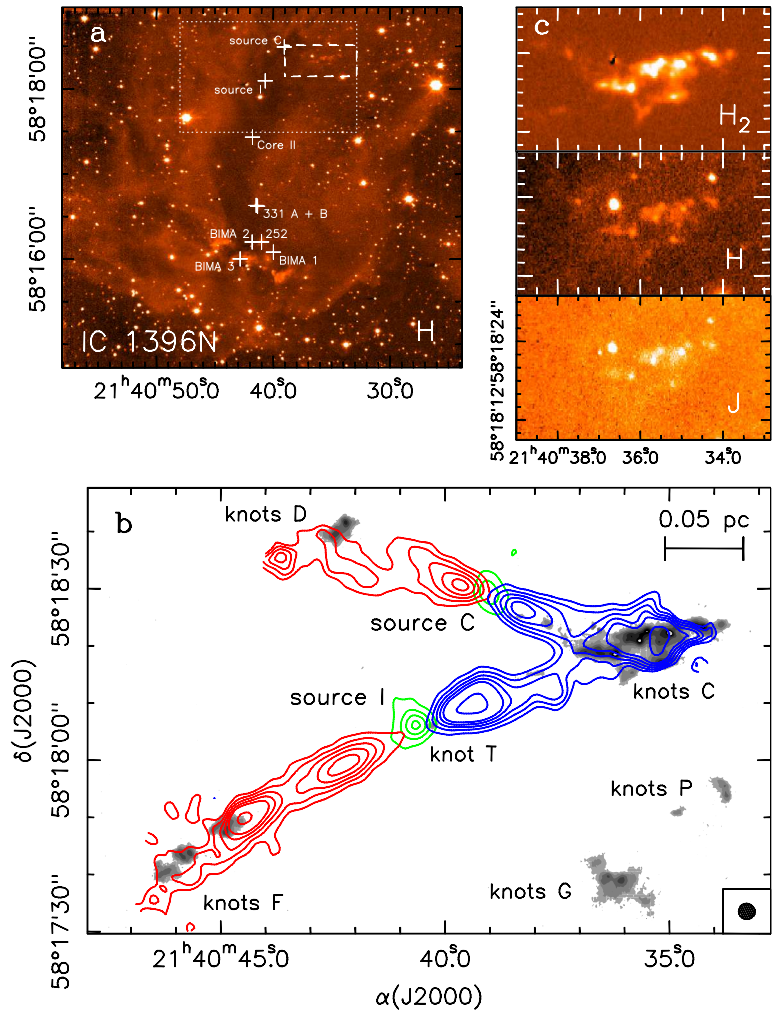}}
\caption{({\it a}) $H$-band image of IC~1396N taken with NICS at TNG. The
white crosses show the positions of the 3.1~mm  sources, BIMA 1, 2,
and 3 from Beltr\'an et al.~(\cite{beltran02}), the Class~I sources \#~252 and
\#~331 (binary, A+B) (Beltr\'an et al.~\cite{beltran09}), the two embedded sources
I and C, which are powering the outflows in panel $b$ (Codella et
al.~\cite{codella01}; this paper), and core~II (Sugitani et
al.~\cite{sugitani02}). The dotted and dashed boxes indicate the region
mapped in panels $b$ and $c$, respectively. ({\it b}) H$_2$ (2.12~$\mu$m) 
image (continuum-subtracted)  in grey-scale (Beltr\'an et
al.~\cite{beltran09}), 2.7~mm continuum emission in green contours  and $^{12}$CO
(\juz) emission averaged over the $[\pm8,\pm38]~\mbox{km s}^{-1}$ velocity
interval in red and  blue contours towards the northern part of
the IC~1396N cloud. Contour levels are 3, 6, 9, 12, 15, and 24 times
1$\sigma$, where 1$\sigma$ is 0.25~mJy\,beam$^{-1}$ for the continuum, and
4~mJy\,beam$^{-1}$ for the CO emission. The PdBI synthesized
beam is shown in the lower right corner. ({\it c}) H$_2$
(continuum-subtracted) ({\it top}), $H$-band ({\it middle}), and $J$-band
({\it bottom}) image towards the chain of knots C (Beltr\'an et
al.~\cite{beltran09}).}
\label{fig1}
\end{figure*}

IC~1396N is a bright-rimmed cloud (BRC38; Sugitani et al.~\cite{sugitani91}), 
located at a distance of 750~pc (Matthews~\cite{matthews79}), which has been
extensively studied by our group (Nisini et al.~\cite{nisini01}; Codella et
al.~\cite{codella01}; Beltr\'an et al.~\cite{beltran02},  \cite{beltran04},
\cite{beltran09}; Fuente et al.~\cite{fuente09}). The cloud shows a cometary
structure elongated in the south-north direction, with the globule head
located at the south, facing the O6.5 exciting star HD~206267, and the tail
pointing to the north. Towards the globule head, one finds the
intermediate-mass protostar IRAS~21391+5802 (BIMA~2), which is associated with
multiple millimeter compact sources and powers a molecular outflow
(Codella et al.~\cite{codella01}; Beltr\'an et al.~\cite{beltran02}; Neri et
al.~\cite{neri07}; Fuente et al.~\cite{fuente09}). Near-infrared (NIR) images
have revealed  a number of small-scale 2.12~$\mu$m H$_2$ line emission
features and Herbig-Haro (HH) flows spread all over the globule (Nisini et
al.~\cite{nisini01}; Sugitani et al.~\cite{sugitani02}; Reipurth et
al.~\cite{reipurth03}; Caratti o Garatti~\cite{caratti06}; Beltr\'an et
al.~\cite{beltran09}), as well as an IR-dark ``tail" that crosses the cloud
from south to north and is clearly pinpointed in the $JHK'$ composite
image (see Fig.~1 of Beltr\'an et al.~\cite{beltran09} and Fig.~\ref{fig1}a).
This IR-dark ``tail" is well-traced in both CO and high-density tracers such as
CS and H$^{13}$CO$^+$ (Codella et al.~\cite{codella01}; Sugitani et
al.~\cite{sugitani02}).

Deep sub-arcsecond 2.12~$\mu$m H$_2$ 1--0~S(1) line emission observations
carried out with the TNG telescope (Beltr\'an et al.~\cite{beltran09}) resolved
the emission features spread throughout the region  into several chains of
knots. In a few cases, these strands of H$_2$ knots have a jet-like morphology,
which suggests that they could trace different flows (Nisini et
al.~\cite{nisini01}; Beltr\'an et al.~\cite{beltran09}). Towards the northern
part of the globule, Nisini et al.~(\cite{nisini01}) proposed that some of these
knots could be associated with the CO molecular outflow discovered by Codella et
al.~(\cite{codella01}), and  Beltr\'an et al.~(\cite{beltran09}) suggested that
strand of knots located at both sides of the IR-dark ``tail"  could be
associated with each other and be part of long flows with lengths ranging from 0.18 to more than
0.4~pc, and that the powering sources were embedded in the ``tail".

To test the hypothesis that some of the molecular hydrogen emission features
could be associated with long flows crossing the northern part of IC~1396N, we
mapped with the IRAM Plateau de Bure Interferometer (PdBI) the whole IR-dark
``tail" in $^{12}$CO, which is the most sensitive outflow tracer and should help
us to trace the cold-warm flow component that is overlooked by the hotter
component tracer H$_2$. The ``tail" was also mapped in the continuum at 2.7~mm
to search for the possible outflow  powering sources that according to Beltr\'an
et al.~(\cite{beltran09}) should be hidden there. In this Letter, we present the
first results of this study, which confirm the scenarios proposed by Beltr\'an
et al.~(\cite{beltran09}), including the collision of two of these flows.

\section{Observations}

We carried out seven-point mosaic observations of the IR-dark ``tail" in
IC~1396N in the 2.7~mm continuum and $^{12}$CO~(\juz) line emission with the
PdBI on 2010 July 24 and 27, and November 4, in the D and C configurations. The
inner hole in the ({\it u, v})-plane has a radius of 7~k$\lambda$.  Line data
were smoothed to a spectral resolution of 1~\kms. The maps were created
with natural weighting, attaining a resolution of $2\farcs9\times2\farcs8$, but
to highlight the continuum and line emission, the final maps were
reconstructed with a circular synthesized beam of $4''\times4''$.  The 1$\sigma$
noise is 0.25~mJy\,beam$^{-1}$ for the continuum map and
10~mJy\,beam$^{-1}$\,channel$^{-1}$ for the CO line map.

\section{Analysis and discussion}

\subsection{The embedded protostars}
\label{protostars}

The continuum observations of the northern part of the IR-dark ``tail" detected
two embedded cores, which we call sources I and C, powering well-collimated
outflows (with collimation factors $>$5 towards the central part of the
outflows; Fig.~\ref{fig1}b) and separated by $\sim$0.1~pc. The source C was
first detected at 1.25~mm with the IRAM 30-m telescope by Codella et
al.~(\cite{codella01}), who suggested that it was the powering source of the
northern bipolar outflow mapped in $^{12}$CO~(\jdu) with an angular resolution
of 10$"$. However, this is the first time that the source is mapped and at high
angular resolution. The existence of source I and the bipolar outflow that is
driving had been suggested by Beltr\'an et al.~(\cite{beltran09}) (see next
section), and our observations here confirm this idea. This new source could be
associated with the H$^{13}$CO$^+$ peak labeled as core I by Sugitani et
al.~(\cite{sugitani02}).  The mass of the sources was estimated assuming a
dust opacity of 1.15~cm$^2$\,g$^{-1}$ at 115~GHz (Beckwith et
al.~\cite{beckwith90}) and a gas-to-dust ratio of 100. Since the dust
temperature, $T_{\rm dust}$, of the sources is unknown, we estimated the
masses for 10~K and 20~K, which are the typical temperatures of the gaseous envelopes
of low-mass protostars. The mass of source I is $\sim$0.5$\,M_\odot$ (for
$T_{\rm dust}$=10~K)  and $\sim$0.2$\,M_\odot$ (for $T_{\rm dust}$=20~K), and
that of source C is $\sim$0.3$\,M_\odot$ (for $T_{\rm dust}$=10~K)  and
$\sim$0.1$\,M_\odot$ (for $T_{\rm dust}$=20~K).

\begin{table*}
\caption[] {Properties of the molecular outflows$^{a,b}$}
\label{tco}
\begin{tabular}{lccccccccc}
\hline
&\multicolumn{1}{c}{$R\,\cos~i$}&
\multicolumn{1}{c}{$t_{\rm out}\,\cot~i$}&
\multicolumn{1}{c}{$T_{\rm ex}$}&
\multicolumn{1}{c}{$M_{\rm out}$}&
\multicolumn{1}{c}{$\dot M_{\rm out}\,\tan~i$}&
\multicolumn{1}{c}{$P^{c}\,\sin~i$}&
\multicolumn{1}{c}{$E^{c}\,\sin^2~i$}&
\multicolumn{1}{c}{$F^{c}\,\sin^2~i/\cos~i$}&
\multicolumn{1}{c}{$L_{\rm mech}\,\sin^3~i/\cos~i$}
\\
\multicolumn{1}{c}{Outflow}&
\multicolumn{1}{c}{(pc)}&
\multicolumn{1}{c}{(yr)}& 
\multicolumn{1}{c}{(K)}&
\multicolumn{1}{c}{($M_\odot$)}&
\multicolumn{1}{c}{($M_\odot$ yr$^{-1}$)}&
\multicolumn{1}{c}{($M_\odot$ \kms)}&
\multicolumn{1}{c}{($10^{43}$ erg)}&
\multicolumn{1}{c}{($M_\odot \, \mbox{km s}^{-1}$\,yr$^{-1}$)}&
\multicolumn{1}{c}{($L_\odot$)}
\\  
\hline
N  Blue       &0.13  &$3.5\times10^3$  &10  &$3.5\times10^{-3}$ &$1.0\times10^{-6}$   &0.06  &1.2  &$1.7\times10^{-5}$  &$0.03$ \\
N  Red        &0.14  &$3.6\times10^3$  &10  &$2.9\times10^{-3}$ &$0.8\times10^{-6}$   &0.05  &0.9  &$1.4\times10^{-5}$  &$0.02$ \\
N  Total      &0.27  &$3.6\times10^3$  &10  &$6.4\times10^{-3}$ &$1.8\times10^{-6}$   &0.11  &2.1  &$3.1\times10^{-5}$  &$0.05$ \\
\hline
S Blue        &0.19  &$4.8\times10^3$  &10 &$5.8\times10^{-3}$  &$1.2\times10^{-6}$   &0.09  &1.7  &$1.9\times10^{-5}$  &$0.03$ \\
S Red         &0.19  &$4.8\times10^3$  &10 &$5.4\times10^{-3}$  &$1.1\times10^{-6}$   &0.10  &2.2  &$2.1\times10^{-5}$  &$0.04$ \\
S Total       &0.38  &$4.8\times10^3$  &10 &$11.2\times10^{-3}$ &$2.3\times10^{-6}$   &0.19  &3.9  &$4.0\times10^{-5}$  &$0.07$\\
\hline
\end{tabular}
 
(a) Estimated values. The outflow parameters should
be corrected if the inclination angle
with respect to the plane of the sky, $i$, is known.  \\
(b) Blueshifted emission averaged over the velocity range [$-38$, $-8$]~\kms\
and redshifted one over [+8, +38]~\kms.  \\
(c) Momenta and kinetic energies are calculated relative to the cloud velocity, $V_{\rm LSR}$=0~\kms.  \\
\end{table*}

\subsection{The colliding molecular outflows}
\label{out}

The elongated structure and the very well-defined bipolar morphology of the two
molecular outflows mapped in $^{12}$CO~(\juz) (Fig.~\ref{fig1}b), suggest that
they are close to the plane of the sky. Both outflows are associated with
2.12~$\mu$m H$_2$  emission. The blueshifted lobes of the two bipolar outflows
strongly overlap towards the position of the H$_2$ strand of knots C.  The
northern outflow, named N, is being powered by source C and is oriented in the
northeast-southwest direction.  Nisini et al.~(\cite{nisini01}) and Beltr\'an et
al.~(\cite{beltran09}) suggested that it could be associated with the strand of
H$_2$ knots D and C. As seen in Fig.~\ref{fig1}b, the CO emission indeed
complements the H$_2$ emission where the excitation conditions and the opacity
of the material are different. On the other hand, the southern outflow, named S,
which is driven by source I and oriented in the southeast-northwest direction,
had never been mapped before. Beltr\'an et al.~(\cite{beltran09}) had proposed
that the chains of H$_2$ knots F, T, and C could be associated with and be part
of the same flow, which would have a length of $\sim$0.4~pc, as the CO
observations have confirmed. These authors had also predicted that the driving
source, in this case source I, should be embedded in the dense gas mapped in CS
by Codella et al.~(\cite{codella01}) and in H$^{13}$CO$^+$ (core I) by Sugitani
et al.~(\cite{sugitani02}). According to the correlations between driving source
and outflow properties of Cabrit and Bertout~(\cite{cabrit92}), the bolometric
luminosity of the powering sources should be $<10$~$L_\odot$. The parameters of
the outflows, which are estimated assuming that the CO emission in the wings is
in  local thermal equilibrium and optically thin, are given in Table~1. The
outflow parameters and the masses of the driving sources are consistent with
those of low-mass protostars (e.g., Arce et al.~\cite{arce07} and references
therein). Since the CO spectra exhibit prominent self-absorption at the cloud
velocity and it is impossible  to derive the excitation temperature, $T_{\rm
ex}$,  from the brightness temperature, the outflow parameters were estimated
assuming $T_{\rm ex}$=10~K, which was the value adopted by Codella et
al.~(\cite{codella01}) for the northern outflow. We used an [H$_2$]/[CO]
abundance ratio of $10^4$ (e.g.\ Scoville et al.~\cite{scoville86}). To
calculate the parameters of the blueshifted lobes, we assumed that half of  the
CO emission seen towards the feature C is associated with outflow N and half
with outflow S. That the parameters of the redshifted lobes are similar
to those of the blueshifted lobes for both outflows suggests that this
assumption is correct. The parameters need to be corrected by the factors indicated
in Table~1 if the inclination angle of the flow with respect to the plane of the
sky, $i$, is known.  On the basis of the morphology of the outflows,
$i\lesssim10\degr$. 

In the scenario proposed by Beltr\'an et al.~(\cite{beltran09}), the outflows N
and S would collide  towards the position of the chain of knots C and this would
explain the strong H$_2$ emission observed towards this feature. As previously
mentioned, the blueshifted lobes of the CO bipolar outflows do indeed overlap
towards the position of the strand of knots C (Fig.~\ref{fig1}b). However, the
collision of outflows is not the only possible explanation of this morphology. 
The overlapping of the blueshifted emission might also indicate that the
outflows are colliding with the same dense clump, or are merely a projection
effect. In the latter scenario, the outflows would be in different planes of the
sky and would be shocking with different dense clumps aligned along the line of
sight. However, that the outflows are almost in the same plane (that of the
sky), that the powering sources are at the same systemic velocity,
\Vlsr$\sim0$~\kms, and that some observational facts favor a collision (see next),
either with the same clump or between outflows, rules out a chance superposition
of the outflows.

\begin{figure}
\centerline{\includegraphics[angle=0,width=8.5cm]{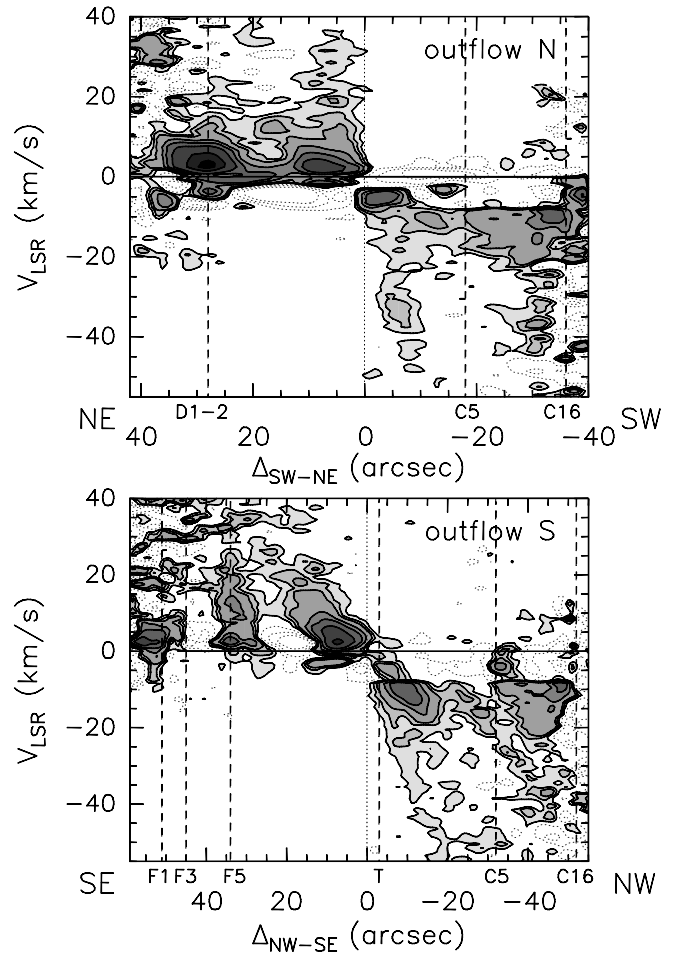}}
\caption{$^{12}$CO~(\juz) position-velocity plots along the major axis of the
outflows, with PA=$-102\degr$ for outflow N, and
PA=$-24\degr$ for outflow S.  The offsets are measured
from the position of source C, $\alpha$(J2000)=$21^{\rm h}40^{\rm h}39\fs04$, 
$\delta$(J2000)=$+58\degr 18'29\farcs8$, for outflow N and of source I,  
$\alpha$(J2000)=$21^{\rm h}40^{\rm h}40\fs66$, $\delta$(J2000)=$+58\degr
18'06\farcs0$,  for outflow S, and are positive towards the northeast and
southeast, respectively. Contour levels are  $-$12, $-$9, $-$3, 3, 6, 9, 12,
27, 45, 90, and 145 times $1\sigma$, where $\sigma$ is 10~mJy\,beam$^{-1}$. The
horizontal line indicates the $V_{\rm LSR}$. The vertical dotted line
indicates the position of the powering sources, and the vertical dashed lines
the position of the H$_2$ knots (Beltr\'an et al.~\cite{beltran09}).}
\label{pv}
\end{figure}

Figure~\ref{pv} shows the position-velocity (PV) plots along the major axis of
the outflows. The cuts were made along PA=$-102\degr$ for outflow N and
PA=$-24\degr$ for outflow S, and all the plots were obtained after averaging the
emission along the direction perpendicular to the cut for the purpose of
increasing the S/N of the plots. The clumpy morphology of the outflows
(Figs.~\ref{fig1}b and \ref{pv}) could be the result of episodic mass-loss
events.  As seen in Fig.~\ref{pv}, the velocity of the CO emission increases
rapidly with increasing distance from the powering sources.This is clearly
visible for the southern outflow. The material  accelerates until it reaches the
position of the H$_2$ knots: the strand of knots D and F, located at the end of
the redshifted lobes of the N and S outflows, respectively,  and the group of
knots C, at the end of the blueshifted lobes. At the position of the H$_2$
knots, there is an increase in the extreme emission velocity
($V$$\gtrsim$40~\kms) for the lowest emission contour level, while the bulk of
the emission abruptly decelerates and has a much lower velocity
($V$$\simeq$10--20~\kms). This is less evident for the knots D, which reach
velocities of only $\sim$20~\kms, although this could be due to the lower
sensitivity at the edge of the map to higher velocities. For the chain of knots C,
the bulk of the emission clearly delineates an inverse triangle shape for both
outflows, with the velocity of the material slowly increasing from the position
of knot C5 (the first knot of the main H$_2$ emission feature encountered by the
blue lobes) to the position where the material at the lowest emission level
reaches the maximum velocity, and then decelerating up to the position of knot
C16 (the last knot of the main H$_2$ emission feature encountered by the blue
lobes). The morphology of the CO emission could be explained in terms of
outflowing material that is impacting, either with a dense clump or with another
outflow, at the position of the H$_2$ knots, giving rise to the NIR emission.

The H$_2$ emission of the group of knots C is the strongest of the entire
northern part of the globule. The integrated line emission of this feature,
summing the emission of all the individual knots, is 
485.4$\times$10$^{-15}$~mW\,m$^{-2}$ (Beltr\'an et al.~\cite{beltran09}). This
value is much higher than the  integrated molecular hydrogen emission of the
strand of knots D (52.8$\times$10$^{-15}$~mW\,m$^{-2}$) and F
(92.9$\times$10$^{-15}$~mW\,m$^{-2}$). This clear asymmetry at both ends of the
outflows, in terms of both morphology and intensity of the feature, led us to think that
the shock that is taking place at the position of the knots C is exceptional. 
The knots D, F, and C are also visible at 4.5~$\mu$m (see Fig.~5 of Beltr\'an et
al.~\cite{beltran09}), in which case, the feature C is also the strongest. We
investigated whether this could be due to extinction. Assuming that knots D and
F have the same integrated emission as knots C but are more extincted, to
reproduce the same integrated fluxes, the visual extinction, $A_{\rm V}$, should
be $\sim$21~mag higher in D than in C, and $\sim$16~mag higher in F than in C.
From the color-magnitude of  Beltr\'an et al.~(\cite{beltran09}), and from
estimates based on the CO and CS emission data of Codella et
al.~(\cite{codella01}), the maximum extinction through the northern end of the
globule ``tail" is $A_{\rm V} < 20$~mag. Although we cannot exclude that the
extinction is severely affecting the intensities of the knots D and F, it seems
unlikely that the difference in extinction between molecular hydrogen features
located outside the IR-dark ``tail", which is the most extincted part of the
globule, are so large. Spectra taken towards the position of knots F and C
(Massi et al.~in preparation), show a richness of other H$_2$ transitions in
$J$-, $H$-, and $K$-bands. The emission is much stronger in C than in F, which
would explain why the strand of knots C is clearly  visible in the $H$ and
$J$-band images (Fig.~\ref{fig1}c).

The CO and H$_2$ emission clearly indicates that there is a shock at  the
position of the strand of knots C. To distinguish between the two possible
scenarios, that is, that the outflows are colliding with each other or are
impacting with the same dense clump, we analyzed the emission at the cloud
velocity traced by  $^{12}$CO in the $[-8,+8]~\mbox{km s}^{-1}$ velocity range
(the cloud \Vlsr\ is $\sim$0~\kms). The average CO emission in this velocity
interval does not show any compact clump at that position. The negative contours
visible in the PV plots at the position of the knots C at systemic velocities
would indicate that the interferometer has filtered out only extended emission.
A compact dense clump is similarly invisible in the channel maps of the
high-density tracer CS~(\jtd) (Codella et al.~\cite{codella01}). Therefore, the
most plausible explanation for the NIR and CO emission is that the outflows are
located in the same plane, as already suggested by their well-defined bipolar
morphology, and are indeed colliding.  This scenario is particularly interesting
because the collision would take place in a very pristine environment, where
sources C and I would be the only embedded protostars. As seen in
Fig.~\ref{fig1}a, most of the protostars are located in the southern part of the
IR-dark ``tail". As far as we know, possible outflow interactions have
previously been reported in the crowded L1448 region, where multiple protostars
and molecular outflows have been detected (Barsony et al.~\cite{barsony98}; Kwon
et al.~\cite{kwon06}). However, the most likely explanation of those
interactions are not outflow collisions but outflow-dense core shocks (Barsony
et al.~\cite{barsony98}).

Assuming that both outflows are randomly oriented, the probability of such an
event is low ($\sim$5\%), if one considers the case of two sources separated by
a short distance (0.2~pc), powering outflows with 0.01~pc of radius, in which
one of the outflows is perpendicular to the line joining the two sources. The
probability is higher ($\sim$10--20\%) if one considers the case in which the
direction of both outflows is allowed to vary. We note that the probability will be even
higher if the separation between the sources is smaller or the opening angle of
the outflow higher. Therefore, that, to our knowledge, clear cases of
outflow collisions have not been found is not because of the rather low probability of
such an event but to the practical difficulty in observing two very close sources in a
similar evolutionary state located at a distance that allows us to resolve the
emission of their respective outflows.

\section{Conclusions}

We have presented evidence for the collision of the outflows N and S observed
in the northern part of the bright-rimmed cloud IC~1396N based on the overlap
of the CO blueshifted lobes, the  well-defined bipolarity of the outflow,
which indicates that both outflows are located close to the plane of the sky,
the strong H$_2$ emission towards the position where the two blueshifted lobes
overlap (strand of knots C), and the non-detection of a compact dense clump at
the position of the putative shock. Additional high-angular resolution
observations of shock tracers, such as SiO, would be needed to confirm this
collisional scenario. Observing lower SiO velocities over the blueshifted
interacting region than for the redshifted emission of the outflow would
further support this scenario. The collision scenario would imply, if
confirmed, that the star formation in this globule is aligned along 
a privileged direction.

\begin{acknowledgements}  This work benefited from research funding from the
European Community's Seventh Framework Programme. 
\end{acknowledgements}

\end{document}